# Photophysics of GaN single photon sources in the visible spectral range

Amanuel M. Berhane,[1] Kwang-Yong Jeong,[2,3] Carlo Bradac,[1] Michael Walsh,[2] Dirk Englund,[2] Milos Toth,[1] and Igor Aharonovich[1]

[1]School of Mathematical and Physical Sciences, University of Technology Sydney, Ultimo,

New South Wales, 2007, Australia.

[2]Department of Electrical Engineering and Computer Science, MIT, Cambridge, MA 02139, USA.

[3]Department of Nano Science and Technology, Gachon University, Gyeonggi-do 13120,

 Republic of Korea

***Abstract:*** *In this work, we present a detailed photophysical analysis of recently-discovered optically stable, single photon emitters (SPEs) in Gallium Nitride (GaN). Temperature-resolved photoluminescence measurements reveal that the emission lines at 4 K are three orders of magnitude broader than the transform-limited widths expected from excited state lifetime measurements. The broadening is ascribed to ultra-fast spectral diffusion. Continuing the photophysics study on several emitters at room temperature (RT), a maximum average brightness of ~(427±215) kCounts/s is measured. Furthermore, by determining the decay rates of emitters undergoing three-level optical transitions, radiative and non-radiative lifetimes are calculated at RT. Finally, polarization measurements from 14 emitters are used to   determine visibility as well as dipole orientation of defect systems within the GaN crystal. Our results underpin some of the fundamental properties of SPE in GaN both at cryogenic and RT, and define the benchmark for future work in GaN-based single-photon technologies.*

## 1. Introduction

Single photon emitters (SPEs) from defects in solids are promising candidates for scalable quantum nanophotonics and nanoscale emitter–cavity systems [1-4]. Among the most studied room-temperature, single-photon emitters, colour centres in diamond [5-9], silicon carbide [10-14], zinc oxide[15-17] and more recently–hexagonal boron nitride [18] are shaping the field. Concurrently, there is a growing push towards identifying and characterizing suitable, new emitters in material systems with well-established growth and nanofabrication protocols. Gallium nitride (GaN) is one such material. For instance, GaN quantum dots (QDs) have been incorporated into nanoscale pillars to generate bright single photons sources in the UV spectral range at cryogenic temperatures and, to some extent, at room temperature (RT)[19-21]. It has also recently been shown that defects in GaN can act as polarized, bright SPEs that operate at RT emitting in the visible  [22, 23] and telecom spectral  range [24]. These SPEs are observed in commercial wafers, which is important for integration with optoelectronic devices and circuits. However, more work is needed to improve our current understanding of the photophysics, atomic structure and technological potential of quantum emitters in GaN. In this work, we perform a detailed photo-physical analysis of optically stable SPEs in GaN. We present cryogenic-temperature spectroscopic data, and a room-temperature analysis of the saturation behaviour and transition kinetics for a range of emitters; we provide detailed polarization measurements at RT.

## 2. Experimental setup

The sample used in this study is a commercially available 4-μm (2-μm p-type/2-μm undoped) GaN crystal grown on sapphire, obtained from Suzhou Nanowin Science and Technology Co., Ltd. The cryogenic-temperature measurements are carried out by placing the sample on a XYZ piezo stage, located in a close-cycle Montana cryostat system with temperature control down to 4 K. All spectroscopy and second-order correlation measurements are carried out using a confocal microscope with an integrated Hanbury-Brown and Twiss (HBT) interferometer. A cw laser (wavelength 532 nm) is used for excitation. The beam is focused to a spot size of ~400 nm through a x100 objective with a numerical aperture (NA) of 0.9. RT polarization spectroscopy is carried out using the confocal microscope where the polarization of the excitation laser is controlled by placing a half-wave plate into the optical path, while a visible polarizer is used in collection to monitor the emission polarization. Quarter-waveplates are used both on the excitation and detection paths to correct for birefringent components. Lifetime measurements are carried out using a pulsed laser (pulse width 32 ps) with a wavelength of 532 nm.

## 3. Results and Discussion

We start by surveying single emitters at cryogenic temperatures. Figure 1(a) shows spectra for 6 emitters selected at random using an excitation laser power of 100 μW, at 4 K. A distinct zero phonon line (ZPL) is obtained for each one of the emitters. The distribution of ZPL peak energy is shown in Figure 1(b) and was obtained by analysing a total of 19 emitters. The ZPL position ranges from 1.736 eV to 1.983 eV with a mean of ~(1.869±0.064) eV. The histogram is similar to that reported for RT GaN (which has a mean of ~1.824 eV) [22], illustrating that the mechanism responsible for the observed ZPL energy range is unaffected by temperature. This is consistent with – and expected from – the interpretation that the energy range corresponds to variations in the positions of individual emitters in cubic inclusions in wurtzite GaN [22]. A histogram of the full width at half maximum (FWHM) of the 19 emitters at 4K is shown in Figure 1(c) with the mean linewidth of ~(3.39±1.12) meV, which is approximately 7 times smaller than the calculated mean linewidth at RT. The additional narrow peak seen at 1.789 eV in Figure 1(a) is the Cr impurity emission from the sapphire substrate.

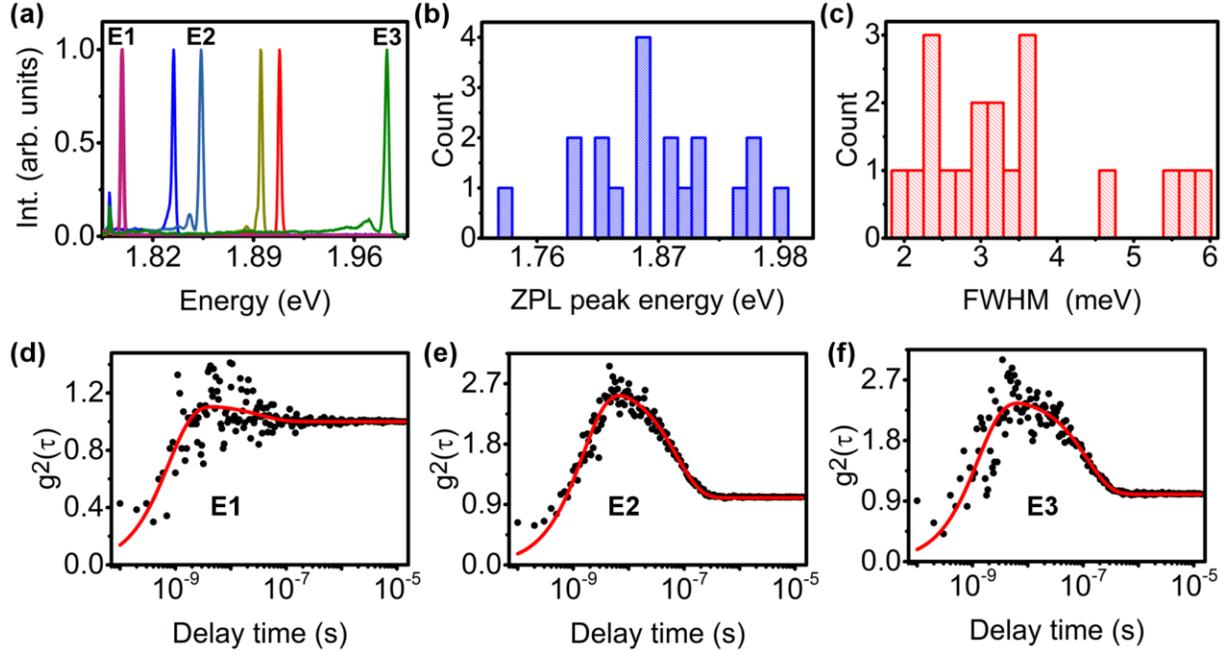

**Figure 1.** Low-temperature (4 K) spectroscopy and photon emission statistics of quantum emitters in GaN. **a)** Representative spectra from 6 emitters with ZPL peak energies of 1.796 eV (E1), 1.834, 1.852 eV (E2), 1.895 eV, 1.908 eV and 1.981 eV (E3). **b)** ZPL peak energy distribution of 19 emitters with mean value of (1.869±0.064) eV. **c)** Histogram showing the FWHM distribution of the emitters in (b) with mean linewidth value of (3.39±1.12) meV. All measurements were taken with an excitation laser power of 100 µW. **d-f)** Second-order autocorrelation measurements for the three emitters labelled E1-E3 in (a) under an excitation power of 50 µW. The curves are fitted with three-level, second-order autocorrelation functions and show that the emitters E1-E3 are single photon emitters with $g^2(\tau=0)$ values of 0.30, 0.27 and 0.18, respectively.

To further characterize the SPEs, we focus on emitters E1, E2, and E3 which approximately span the observed ZPL range, as shown in Figure 1(a), and have ZPLs' FWHM of 1.796 eV (1.6 meV), 1.852 eV (2.4 meV) and 1.981 eV (2.3 meV), respectively. The emitters are photostable, as is illustrated by the singlephoton fluorescence time-traces shown in Figure S1. Figure 1(d-f) shows second-order autocorrelation measurements for correlation times of to up to 15 µs. The single-photon nature of the light emitted from E1-E3 is revealed by $g^2(\tau=0)$ values of 0.30, 0.27 and 0.18, respectively. The deviations from zero are due to background fluorescence and detector jitter. The data can be fitted optimally by a three-level model with a long-lived metastable state:

$$g^2(\tau) = 1 - (1+a)e^{|\tau|/|\tau_1|} + ae^{|\tau|/|\tau_2|} \quad (1)$$

where $a$, $\tau_1$ and $\tau_2$ are the scaling factor for bunching, excited state lifetime and metastable state lifetime, respectively. The values of $a$, $\tau_1$ and $\tau_2$ obtained by fitting $g^2(\tau)$ are summarized in Table 1. Although the bunching behaviour (seen as a peak in each $g^2(\tau)$ profile) is different for each one of the emitters, it is clear that a shelving state observed at RT [22] persists at cryogenic temperatures. Extended photon correlation measurements of up to 0.1 s (Fig. S2) confirm the absence of additional, longer-lived metastable states, with the $g^2(\tau)$ profiles remaining constant to up to 0.1 seconds.

|  | E1 | E2 | E3 |
|---|---|---|---|
| $a$ | 1.44 | 1.71 | 0.12 |
| $\tau_1$ (ns) | 1.29 | 1.55 | 0.76 |
| $\tau_2$ (ns) | 118 | 73.2 | 35.3 |

**Table 1.** Parametric values $a$, $\tau_1$ and $\tau_2$ obtained by fitting the second-order autocorrelation functions of E1, E2 and E3, assuming three-level system dynamics.

To obtain lifetime information from emitters E1-E3, time-resolved PL measurements were carried out using a 532-nm pulsed excitation laser (Fig. 2(a-c)). Double-exponential fits (red lines) of the experimental data yielded lifetimes of 1.6 ns, 2.7 ns and 2.0 ns (where the fast decay component of each fit corresponds to the system response). Based on the measured lifetimes, the calculated Fourier transform limited linewidths, $\Gamma$, of emitters E1, E2 and E3 are 0.4, 0.2 and 0.3 µeV, respectively. These values are roughly three orders of magnitude lower than the measured linewidths presented above for the three emitters. A similar, large deviation in the natural linewidth has been reported previously in off-resonantly excited single GaN and InGaN quantum dots (QDs), and it is associated with spectral diffusion[25, 26]. The major cause for spectral diffusion in QDs is charge fluctuations, which are likely exacerbated in GaN by the large in-built electric field caused by the spontaneous and piezoelectric polarization of GaN [25, 27]. Hence, we attribute the ZPL broadening seen in Fig. 1(c) to ultrafast spectral diffusion (at µs- or ns-time scales). Ultrafast spectral diffusion has been reported as a mechanism for broadening in other materials such as SiC and diamond [12, 28].

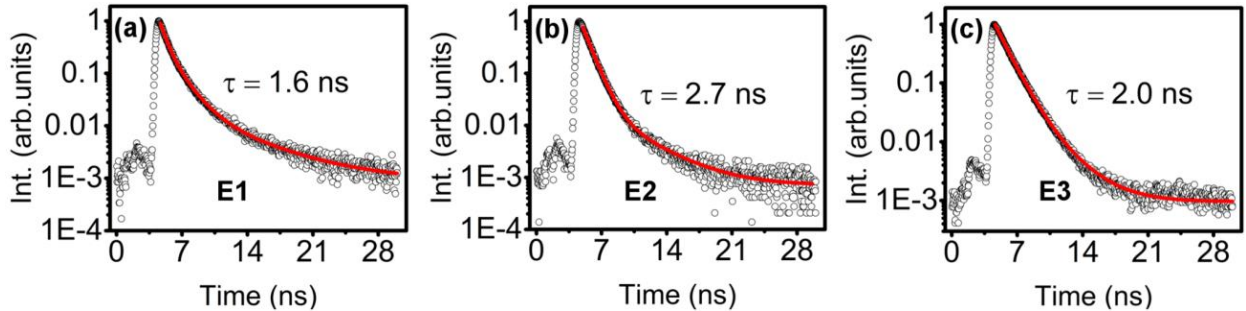

**Figure 2.** PL decay time measurements of quantum emitters in GaN obtained at 4 K using a 532-nm pulsed excitation laser. **a-c)**. Double exponential fits (red line) of the background-corrected measurements yield excited state lifetimes of 1.6, 2.7 and 2.0 ns for emitters E1, E2 and E3, respectively.

To gain more information on ZPL stability, we performed time-resolved spectroscopy (Fig. 3(a-c)). The mean ZPL peak position for emitters E1-E3, at an excitation power of 50 µW, is (1.796±0.0002) eV, (1.852±0.0005) eV and (1.981±0.0002) eV, showing that the ZPLs are stable and there is no substantial spectral diffusion at a time-scale of seconds. Thus, faster (µs to ns) measurements are required to probe the aforementioned ultrafast spectral kinetics.

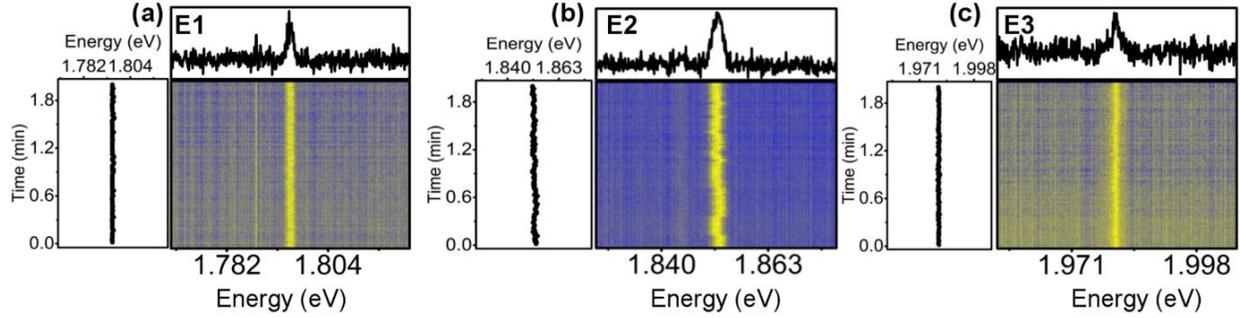

**Figure 3.** Time-resolved PL spectra of the emitters E1-E3 obtained at 4 K using an excitation power of 50µW. **a-c)** ZPL peak energy (left) measured every second for 2 minutes. The spectral maps show the yellow bright points as the peaks of the ZPL corresponding to the integrated spectrum (top) for each emitter. A stable mean ZPL peak energy of (1.796±0.0002) eV, (1.852±0.0005) eV and (1.981±0.0002) eV is observed for E1, E2 and E3, respectively.

Figure S3(a-c) shows the ZPL shift ($\Delta E_{ZPL}(T)$) for the three emitters as a function of temperature, where the shift is calculated as $\Delta E_{ZPL}(T) = E_{ZPL}(T) - E_{ZPL}(4K)$. It has been shown previously that the ZPL energy of emitters in GaN exhibit an unusual S-shape (inverted S-shape) dependence on temperature which indicates that the defects are located in cubic inclusions in wurtzite GaN[22]. Here, the same behaviour is obtained for the three emitters E1-E3 (Figure S3(a-c)). Also, consistent with the previous study, the temperature-dependent broadening of the FWHM for E1-E3 shown in Figure S3(d-f) deviates from the monotonic temperature dependence that is typical of simple defect systems[12, 29, 30].

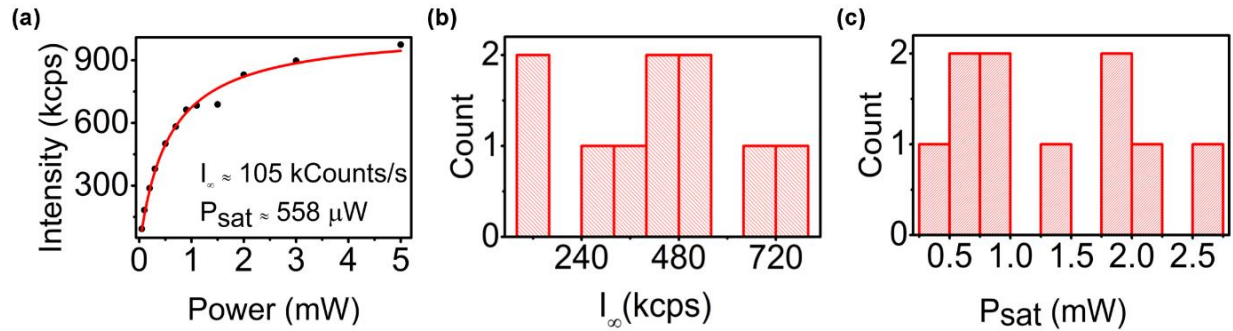

**Figure 4.** Saturation behaviours of emitters in GaN at room temperature. **a)** Background-corrected fluorescence intensity versus power from a representative emitter with a ZPL at 1.818 eV, and a maximum intensity of ~105 kCounnts/s at a saturation power of 558 µW. **b, c)** Statistical distribution of the maximum intensity and saturation power from 8 emitters, with a mean value of ~(427±215) kCounts/s and ~(1270±735) µW, respectively.

Next, we investigate the photophysics of multiple emitters at RT. We measured the brightness of 9 SPEs and extracted the maximum fluorescence intensity of each emitter. Figure 4(a) shows an example of background-corrected, power-dependent saturation behaviour for a representative emitter in GaN. It has a RT ZPL at 1.818 eV for optimized absorption polarization direction. The power-dependent emission intensities are fitted using the relationship:

$$I = I_\infty \frac{P}{P + P_{sat}} \qquad (2)$$

where $I_\infty$ is the maximum intensity and $P_{sat}$ is the saturation power, yielding ~105 kCounts/s and ~558 µW, for the representative emitter. The histograms in Figure 4(b-c) summarize the maximum intensity and saturation power for 8 additional emitters. The mean value of maximum intensity, $I_\infty$, is ~(427±215) kCounts/s, where all emitters are excited using a 532-nm cw laser. This is comparable to other emitters in 3D crystals and can be further improved by employing solid immersion lenses or nanoscale pillars[13, 31-33]. The considerable standard deviation observed in the mean fluorescence intensity, as well as the saturation power, ~(1270±735) µW, may arise from discrepancies in the effective absorption of the off-resonant excitation by the different emitters due to their multiple orientations in the film, as is discussed below[34].

Given the variability in brightness of different SPEs in GaN, we investigated the power-dependent antibunching characteristics of individual emitters, at different fractions of their respective saturation powers ($P_{sat}$). Figure S4(a-c) shows power-dependent, second-order autocorrelation functions for the 3 emitters E1-E3. The measurements reveal bunching statistics at intermediate time scales for increasing excitation powers, confirming the involvement of shelving states in the transition kinetics at RT. The power dependent antibunching characteristics are well fitted using a second-order autocorrelation function accounting for a three-level system (equation S1), i.e. Eq. 1 re-written in terms of decay rates. The antibunching ($\lambda_1$) and bunching ($\lambda_2$) decay rates, as well as the scaling factor for the bunching ($a$) are determined for each one of the 3 emitters, at different powers. The strength of the bunching behaviour at different values of $P_{sat}$ varies between emitters, as expected from the large differences in saturation behaviours discussed above.

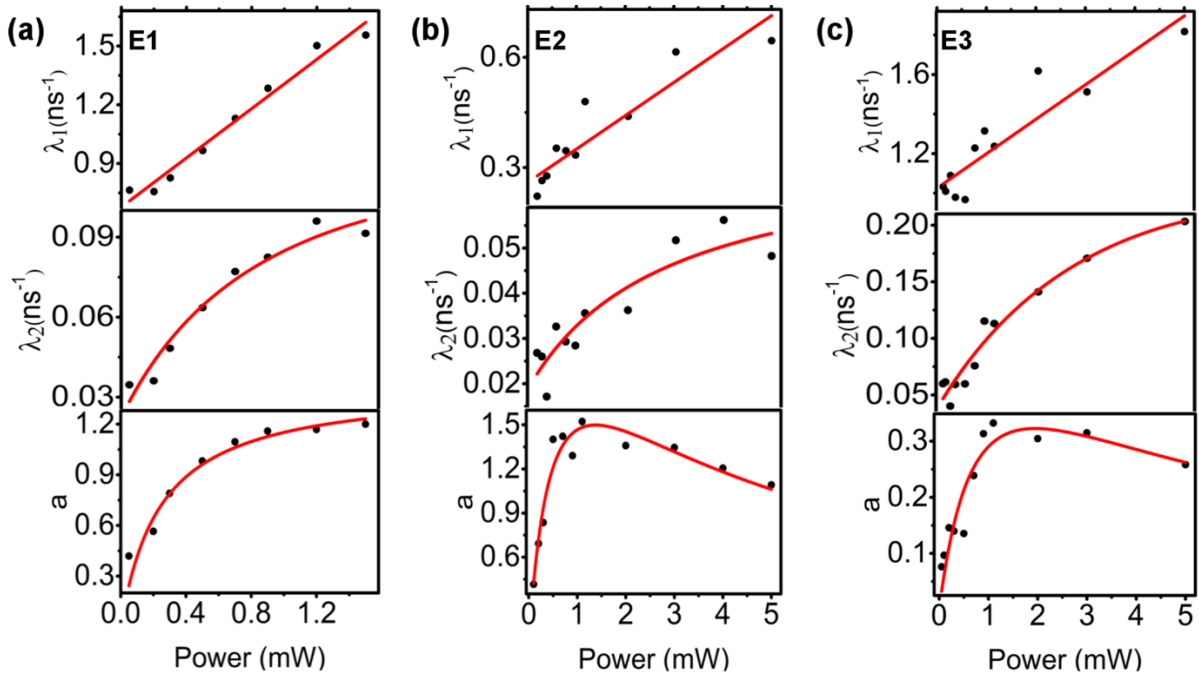

**Figure 5.** Power-dependent properties of the decay rates $\lambda_1$, $\lambda_2$ and the scaling factor $a$ for the 3 emitters E1-E3 measured at room temperature. The data points (black dots) are fitted well (red lines) by considering three-level transition kinetics.

Fitting the power-dependent behaviours of $\lambda_1$, $\lambda_2$ and $a$ for each emitter in Figure 5(a-c) via a three-level transition kinetics model, we determine the characteristic rate-coefficients and relevant parameters (see supplementary information). Table 2 summarizes these values for each emitter together with their corresponding ZPL energy. The non-radiative decay in E2 and E3 occurs via a power-dependent shelving mechanism with a positive value for $\beta$ (See supplementary information). On the other hand, the non-radiative decay in E1 occurs via a power-independent shelving state where $\beta = 0$. The difference in the value of $\beta$ indicates a varying dependence of absorption cross-section for the individual shelving state, with varying excitation power.

| SPE | ZPL (eV) | $\kappa_{21}$ (ns$^{-1}$) | $\kappa_{23}$ (ns$^{-1}$) | $\kappa_{31}$ (ns$^{-1}$) | $\kappa_{31}^0$ (ns$^{-1}$) | $\alpha$ ($\mu W^{-1}$) | $\beta$ ($\mu W^{-1}$) |
|---|---|---|---|---|---|---|---|
| E1 | 1.934 | 0.678 | 0.127 | 0.024 | - | 9.27x10$^{-4}$ | 0 |
| E2 | 1.818 | 0.268 | 0.046 | - | 0.021 | 3.31x10$^{-4}$ | 3.29x10$^{-5}$ |
| E3 | 1.826 | 1.039 | 0.521 | - | 0.043 | 1.65x10$^{-4}$ | 3.49x10$^{-4}$ |

**Table 2.** Rate coefficients extracted for the selected three SPEs by fitting their power-dependent parameters in Figure S4. The quantities $\kappa_{12}$, $\kappa_{21}$, $\kappa_{23}$ and $\kappa_{31}$ are the rate coefficients for transitions between coupled states $|1\rangle \to |2\rangle$, $|2\rangle \to |1\rangle$, $|2\rangle \to |3\rangle$ and $|3\rangle \to |1\rangle$, respectively (see supplementary information). All emitters except E1 show a power-dependent shelving state. α and β are linear fitting parameters for the power dependence of $\kappa_{12}$ and $\kappa_{31}$, respectively.

The excited state lifetime, $\tau_{|2\rangle}$, is calculated for the three emitters using the expression $\tau_{|2\rangle} = (\kappa_{21} + \kappa_{23})^{-1}$ [35]. As shown in Table 3, emitters in GaN have short lifetimes with moderate variability between them. This is consistent with the aforementioned lifetime values of emitters in GaN obtained with pulsed-laser excitation. Also, $\kappa_{21}$ is ~2–40 times larger than $\kappa_{23}$, indicating the strong propensity of the excited state to decay radiatively to the ground state, rather than via the 'dark' shelving state. One of the advantages of the rate analysis, using this approach, is that it allows us to estimate the metastable lifetime ($\tau_{|3\rangle}$) separately from the excited state lifetime. The quantity $\tau_{|3\rangle}$ is given by $1/\kappa_{31}$ or $1/\kappa_{31}^0$ in the case of a power-dependent shelving state[35].

|  | E1 | E2 | E3 |
|---|---|---|---|
| $\tau_{|2\rangle}$ (ns) | 1.2 | 3.2 | 0.6 |
| $\tau_{|3\rangle}$ (ns) | 41.7 | 47.6 | 23.3 |

**Table 3.** Calculated values of excited state ($\tau_{|2\rangle}$) and metastable state lifetime ($\tau_{|3\rangle}$) for the three emitters

Finally, we focus on the polarization behaviour of emitters at RT. Figure 6(a) shows a polar plot from a representative emitter with a RT ZPL at 1.818 eV. The absorption polarization [green] profile is traced using a half-wave plate to highlight the angle at which the minimum and maximum intensities occur. By fixing the half-wave plate at an angle where maximum absorption intensity occurs, and rotating the visible polarizer, we obtain the emission polarization profile for the representative emitters [red] in Figure 6(a). The polarization data are fitted with the function $I(\phi) = a + b\cos^2(\phi)$ where $a$, $b$ and $\phi$ are offset parameter, initial intensity amplitude and angle between excitation and dipole orientation, respectively[36]. Fitting the emission polarization with this function, we determine the minimum ($I_{min} = I(\phi = 90°) = a$) and maximum ($I_{max} = I(\phi = 0) = a + b$) of the emission (absorption) polarization direction for the representative emitter in Figure 6(a) to occur at ~135° (60°) and ~40° (140°), respectively. Such analysis of polar plots is useful as it allows for the easy determination of the dipole polarization visibility, as well as the relative orientations of the absorption and emission polarization for individual emitters [37-39].

The polarization visibility is given by the intensity contrast equation:

$$I = \frac{I_{max} - I_{min}}{I_{max} + I_{min}} \quad (3)$$

which, with regard to the emitter shown in Figure 6(a), yields values for the absorption and emission polarization visibility of 34% and 79%. Notably the absorption is not fully polarized, while the emission is. We carried out similar measurements on 14 additional emitters. Figure 6(b) shows the histogram of their percentile visibility in absorption (green) and emission (red). Interestingly, we observe a significant discrepancy in absorption polarization visibility, which ranges between 26% and 94% with the mean value at~(57±26)%. The rather large distribution in absorption dipole visibility, results in the large distribution of laser power required to saturate the emitters, as discussed before. This significant difference in excitation visibility as well as saturation behaviour indicates that the orientation of the dipole varies from emitter to emitter, with weak visibility being likely the result of the dipole having significant out-of-plane components in the 3D crystal[36, 39].

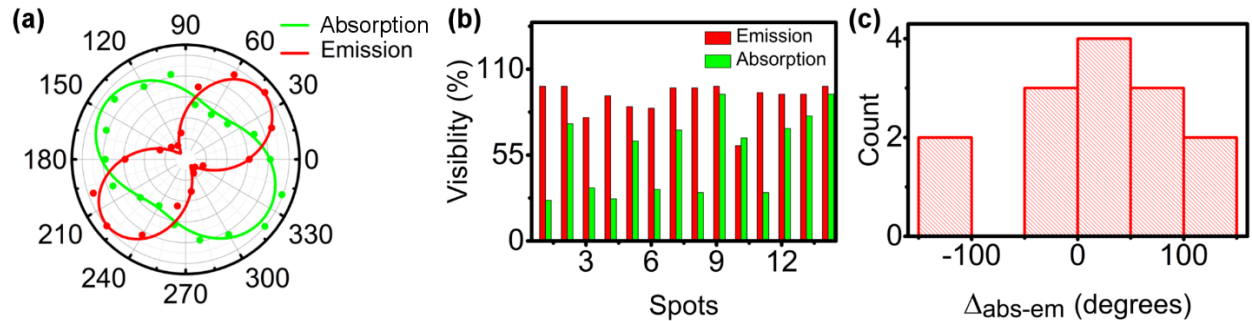

**Figure 6.** Room-temperature polarization spectroscopy of emitters in GaN. **a)** Absorption (green) and emission (red) polarization profiles from an emitter with a ZPL at 1.818 eV, exhibiting polarization visibilities of 34% and 79%, respectively. **b)** Polarization visibilities of 14 emitters showing that while the emitters are strongly polarized in emission, they show variable degrees of absorption polarization. **c)** Histogram of the difference in orientation between absorption and emission polarization.

On the other hand, a mean emission polarization visibility of ~(91±11)% is obtained for the same emitters. This visibility is a strong indicator that these emitters in GaN are linearly polarized (ideal case ~100% for a single dipole). We attribute the deviation from the ideal value of the visibility to fluorescence aberrations arising from residual birefringence and imaging through a high-NA (0.9) objective [36, 40].

The relative orientation between the absorption and the emission dipoles is further analysed for the 14 emitters showing a misalignment ranging from -110° to 120° as shown by the histogram in Figure 6(c). This is expected, considering that the emitters are believed to be point defects located in cubic inclusions. Consequently, for off-resonant excitation, absorption may involve a transition to an excited state of the inclusion. The emission transition, however, involves only the highly localized levels of the defect thus, giving a more distinct radial emission direction compared to that of absorption.

To elucidate the nature of the preferential excitation axis, the maximum absorption polarization angle is measured for the 14 emitters and compared to the wurtzite crystal plane directions in Fig. 7. The angle spans all directions, with a maximum occurrence at ~140°, which corresponds to the $[1\bar{1}00]$ lattice direction of the (0001) wurtzite GaN, as shown in Figure7(b). The rotational orientation of the sample was deduced by considering that the sample is mounted with the unit cell along the (0001) plane almost parallel to the excitation field. In this arrangement, rotating the excitation polarization through 180° sweeps all planes of the hexagonal unit cell. Hence,the highest occurrence angle of 140° which corresponds to the lattice plane direction $[1\bar{1}00]$ is believed to contain the greatest density of cubic inclusions. Furthermore, the highest absorption visibility of 94%, is observed from an emitter with an orientation angle of ~16°, which corresponds to a minimal out-of-plane orientation. This is expected since confined exciton separation occurs along the c axis, where the excitation is aligned parallel to the c axis[41].

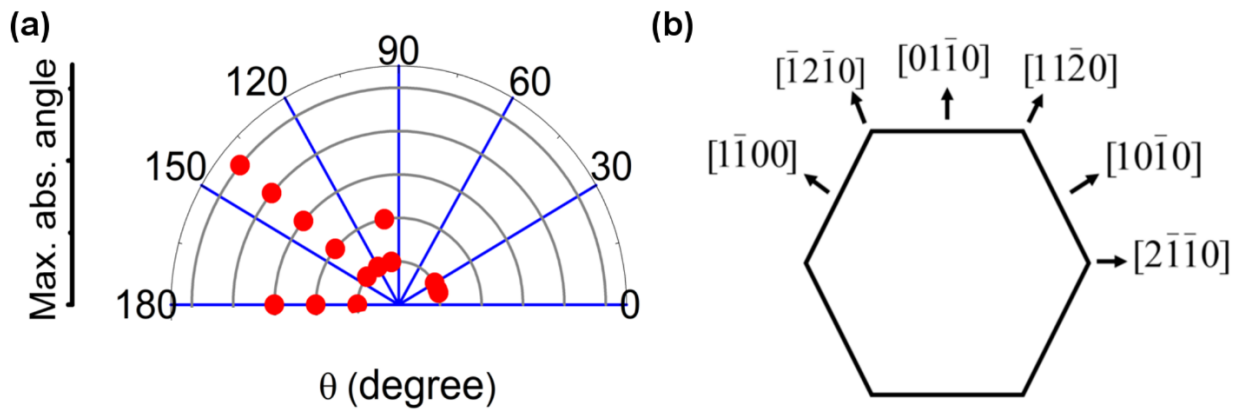

**Figure 7.** Maximum absorption angles for the 14 emitters shown in Fig. 6. **a)** Scatter plot of the maximum absorption axis of the 14 emitters. **b)** Fundamental lattice directions of the wurtzite unit cell, showing that the maximum in angular distribution in (a) corresponds to the $[1\bar{1}00]$ lattice direction of wurtzite GaN.

## 4. Conclusion

To conclude, we carried out low-temperature spectroscopy of SPEs in GaN. While the FWHM of GaN emitters is significantly narrower at 4 K than at RT, the FWHM does not approach the Fourier Transform limited linewidths. Ultrafast spectral diffusion is the most likely explanation for the line broadening where future experimental investigation using approaches such as correlation interferometery should confirm this hypothesis[28]. Temperature-dependent ZPL shift and FWHM broadening results confirm the previously proposed explanation for the existence of a cubic inclusion near the radiative point defect as the cause for the observed S-shaped ZPL shift and non-monotonic broadening.

A saturation behaviour of the emitters was also measured, showing brightness difference among emitters with an average saturation count rate of ~(427±215) kCounts/s. The emitters' kinetics can be described efficiently using a three-level system. Polarization measurements from multiple emitters show high emission visibility of more than 90% and varying strength in the absorption cross section under excitation with a linearly polarized, off-resonant laser. This work evaluates SPEs in GaN as strong alternatives for application in quantum technologies; at the same time, it highlights bottlenecks hindering their immediate implementation.

**Acknowledgments**

This work was supported in part by the U.S. Army Research Laboratory (ARL)(FA9550-14-1-0052) and the Air Force Office of Scientific Research (AFOSR) (FA9550-14-1-0052). Financial support from the Australian Research council (via Grant Nos. DP140102721), FEI Company, and the Asian Office of Aerospace Research and Development Grant No. FA2386-15-1-4044 are gratefully acknowledged